\documentclass[twocolumn]{aastex631}  
\usepackage{graphicx,times,epsf}            
\usepackage{longtable}
\usepackage{natbib}
\usepackage{amssymb}
\usepackage{amsmath}
\usepackage{soul,xcolor}
\shortauthors{Yang et al.}
\begin{document}

\title{Tentative Evidence for Transit Timing Variations of WASP-161b}
\email{sailoryf@nao.cas.cn; sailoryf1222@gmail.com}
\email{rchary@ipac.caltech.edu}
\author[0000-0002-6039-8212]{fan yang}
%\affil{IPAC, Caltech, KS 314-6, Pasadena, CA 91125, USA\\}
\affil{National Astronomical Observatories, Chinese Academy of Sciences, 20A
Datun Road, Chaoyang District, Beijing 100101, China\\}
\affil{Department of Astronomy, Beijing Normal University, Beijing 100875,
People's Republic of China\\}
\affil{School of Astronomy and Space Science, University of Chinese Academy of Sciences,
Beijing 100049, China\\}

\author[0000-0001-7583-0621]{ranga-ram chary}
\affil{Caltech/IPAC, KS 314-6, Pasadena, CA 91125, USA\\}

\begin{abstract}
We report on the detection of transit timing variations (TTV) of WASP-161b by using the combination of TESS data and archival data. The midpoint of the transits in TESS data are offset by $\sim$67 minutes in Jan. 2019, and $\sim$203 minutes in Jan. 2021, based on the ephemeris published in previous work. We are able to reproduce the transit timings from the archival light curve (SSO-Europa; Jan. 2018) and find that the timing is consistent with the published ephemeris under a constant period assumption. Conversely, we find that the transit midpoint of the SSO-Europa light curve indicates a 6.97-minute variation at 4.63 $\sigma$ compared to the prediction obtained from TESS timings, and a constant orbit period assumption. The TTVs could be modeled with a quadratic function, yielding a constant period change. The period derivative $\dot{P}$ is -1.16$\times$10$^{-7}\pm$2.25$\times$10$^{-8}$ days per day (or $-3.65$ s/year), using timings obtained from SSO-Europa and TESS light curves. Different scenarios, including a decaying period and apsidal precession, can potentially explain these TTVs but they both introduce certain inconsistencies.
%We find that tidal dissipation in the planet can explain the timing variations which provides a unique pathway toward studying the internal structure of the exoplanet. 
%We have obtained CHEOPS observations for two transits in Jan. 2022 to distinguish between different TTV scenarios. We expect the timing to vary by 5 minutes, compared to the timing predicted from SSO-Europa and TESS with a constant period assumption.
\end{abstract}
\keywords{Exoplanet systems (484), Exoplanet astronomy (486), Transit photometry(1709), transit timing variation method (1710)}

\section{Introduction}

The transit timing variations (TTVs) of hot Jupiters could potentially leave an imprint of multiple physical processes such as tidal dissipation, apsidal precession, R$\o$mer effect, and mass loss \citep{Ragozzine2009, Valsecchi2015, 2017wasp12b}. Direct detection of TTVs requires a long observational baseline and high precision in identifying the transit midpoint. This can be a challenge due to bandpass mismatches, light curve sampling systematics, and lack of continuity of observations through transit ingress and egress. The launch of the Transiting Exoplanet Survey Satellite \citep[TESS;][]{Ricker2015} in 2018, together with the {\it Kepler} mission \citep{Kepler2010} launched in 2009 and subsequent K2 mission \citep{K2}, provide long time baselines when combining the timing measurements with ground-based telescopes \citep{wasp2006,kelt2007,IvshinaJosh}. The long timing baselines and high precision transit measurements enable one to assess the presence of TTVs with great fidelity.
For example, WASP-12b has been reported to show an orbit decaying with time and possibly with hints of precession \citep{2011wasp12b,2012wasp12b,2017wasp12b,2020wasp12b, 2021wasp12bTESS}. WASP-4b has recently been reported to show a transit which is earlier than expected by 81.6$\pm$11.7 s \citep{WASP-4b} in TESS data.

Monitoring of TTVs could potentially distinguish between different scenarios for the origin of hot Jupiters \citep{Agol2005, Dawson2018}. The two favored scenarios are migrations due to dynamical scattering \citep{Rasio1996, Wu2003, Lendl2014}, and migration due to dynamical friction between the circumstellar disk and the planet \citep{Goldreich1980, Lin1996}. If TTVs show the presence of
short-period planets with high eccentricity ($e$), that would favor the former scenario since the circularization of the orbit due to tidal dissipation should happen rapidly \citep{circularization}. In the latter case, the orbits of the hot Jupiters should be circularized on the timescale of the lifetime of the disk. 
%TTVs in short period planets with high $e$ are very rare because any high $e$ should be very quickly circularized.

The giant planet WASP-161b revolves around a bright F6-type star (10.98 mag, V band) with an orbital period of 5.41 days \citep{discoverpaper}. The host star has an $R_{\ast}$ of $1.71\pm0.08R_{\odot}$, an $M_{\ast}$ of 1.39$\pm$0.14$M_{\odot}$, while the planet has an $R_{P}$ of $1.14\pm0.06R_{J}$, and an $M_{P}$ of 2.49$\pm$0.21$M_{J}$.

We report here on the detection of TTVs of WASP-161b through a combined analysis of TESS data taken in 2019 and 2021, with archival data from 2018 \citep{discoverpaper}. The paper is organized as follows. In Section 2, the data reduction and timing deriving are described. In Section 3, we present the timing deviation of WASP-161b and possible physical explanations for the deviation. In Section 4, we summarize our conclusions.

\section{Data Reduction and Timing Derivation}

The ephemeris has been reported in the discovery paper \citep{discoverpaper} while we derived transit parameters from TESS observations in 2019 and 2021. We also re-analyzed the archival light curves from the discovery paper \citep{discoverpaper}. Specifically, we find only the light curves from 2017 (TRAPPIST-North) and 2018 (SSO-Europa) are available. We note that SSO-Europa light curve is the only archival light curve that covers the whole transit window, all other archival light curves cover only partial transits. Detrending of partial transits is fraught with uncertainty; we therefore ignore data from partial transits. SSO-Europa light curve has a non-uniform sampling of $\sim$ 20s with an exposure time of 10s. 
All the available timings are listed in Table \ref{table: parameters}.

\subsection{TESS Data and Light Curve Generation}

TESS takes exposures of 2 seconds duration but co-adds the data into science products of three different cadences, due to data downlink limitations. The individual frames in the vicinity of sources of high interest are released with a cadence of 2 minutes, namely Target Pixel Files (TPF). The 30-minute cadence data for all the sources in the field of view are available as Full Frame Images \citep[FFI;][]{Ricker2015}. 10-minute cadence FFI is available for the extended mission data. WASP-161b has been observed in TESS Sector 7 (Jan. 2019) with a 30-minute cadence and Sector 34 (Jan. 2021) with 2-minute and 10-minute cadence (data available in MAST: \dataset[10.17909/t9-yk4w-zc73, 10.17909/t9-nmc8-f686].).

We developed a photometric pipeline to obtain light curves from TESS images. The pipeline includes modules for astrometry correction, nearby source deblending, circular aperture photometry, and light curve detrending. The details of the pipelines are described in \citet{Yangatmos}. The light curves obtained from the pipeline are consistent with light curves generated by TESS Pre-search Data Conditioning (PDC) \citep{PDC} among the comparison sample of objects \citep{Yangatmos, YangLD}. In this work, the fitting results from our pipeline are consistent within 1 $\sigma$ to that derived from the PDC light curves for the 2021 data and Quick Look Pipeline (QLP) light curves \citep{Twicken2016, Huang2020} for the 2019 data. As a result, all the parameters quoted in the paper are using our pipeline.

\begin{figure}
  \centering
   \includegraphics[width=4in]{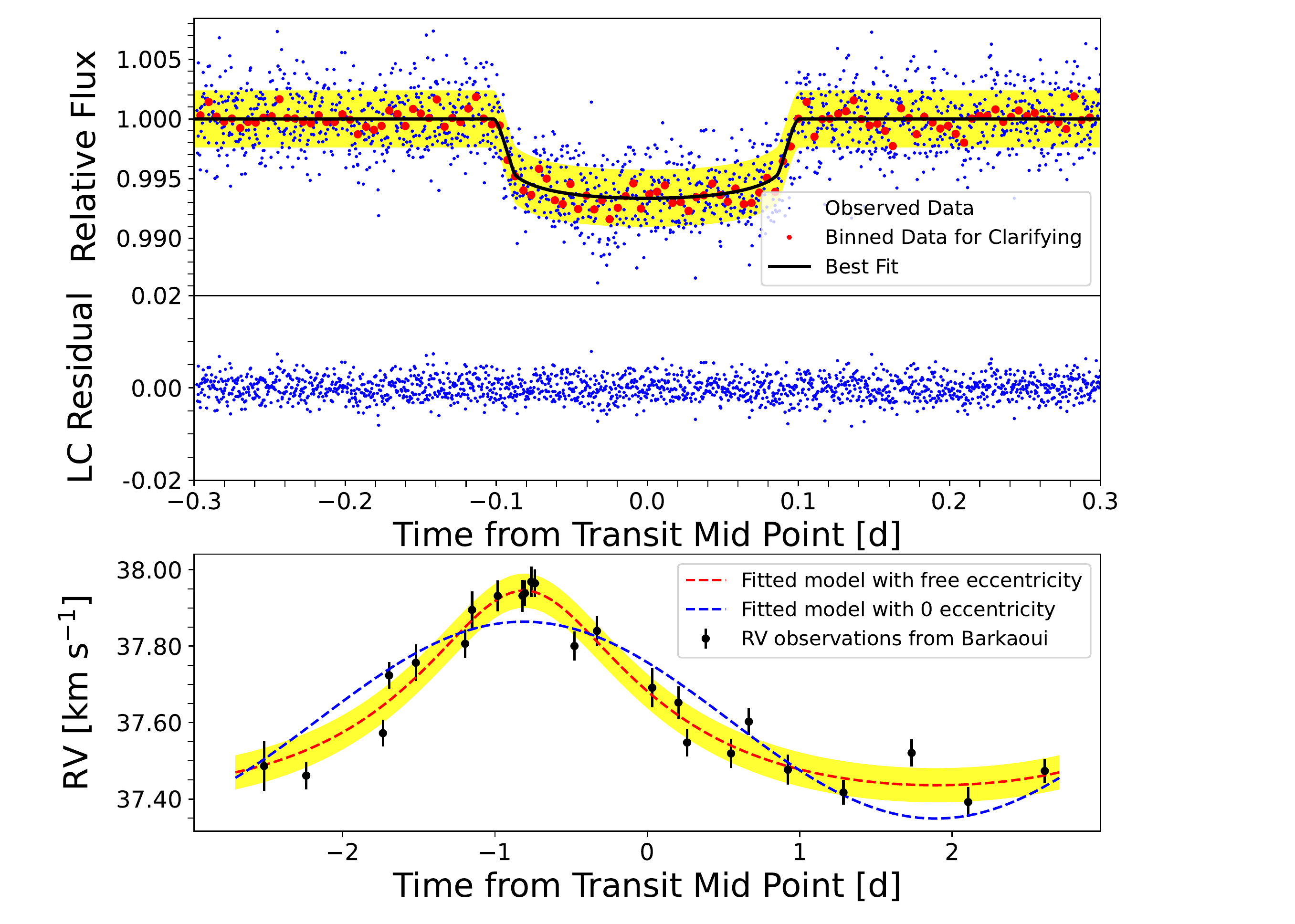}
  \caption{Combined-model fit to TESS light curve (LC) data, and the radial velocity (RV) data from \citet{discoverpaper}. Upper panel: the blue points are photometry from 2021 observations; red points, binned observations for clarity; black line, fitted model to the light curve; yellow region, 1 $\sigma$ significance region encompassing 68\% of the data points. The fitting residual is given in the middle panel as blue points. The p-value from the Shapiro-Wilk test is 0.223, indicating a good Gaussian distribution. Lower panel: the black points indicate the RV observations. The red dash line gives the RV fitting with free eccentricity. The blue dash line shows the RV fitting with 0 eccentricity - it clearly is a worse fit to the data as quantified in Table 1. 
  }
\label{image: combinedfit} 
\end{figure}

\begin{figure}
  \centering
   \includegraphics[width=4in]{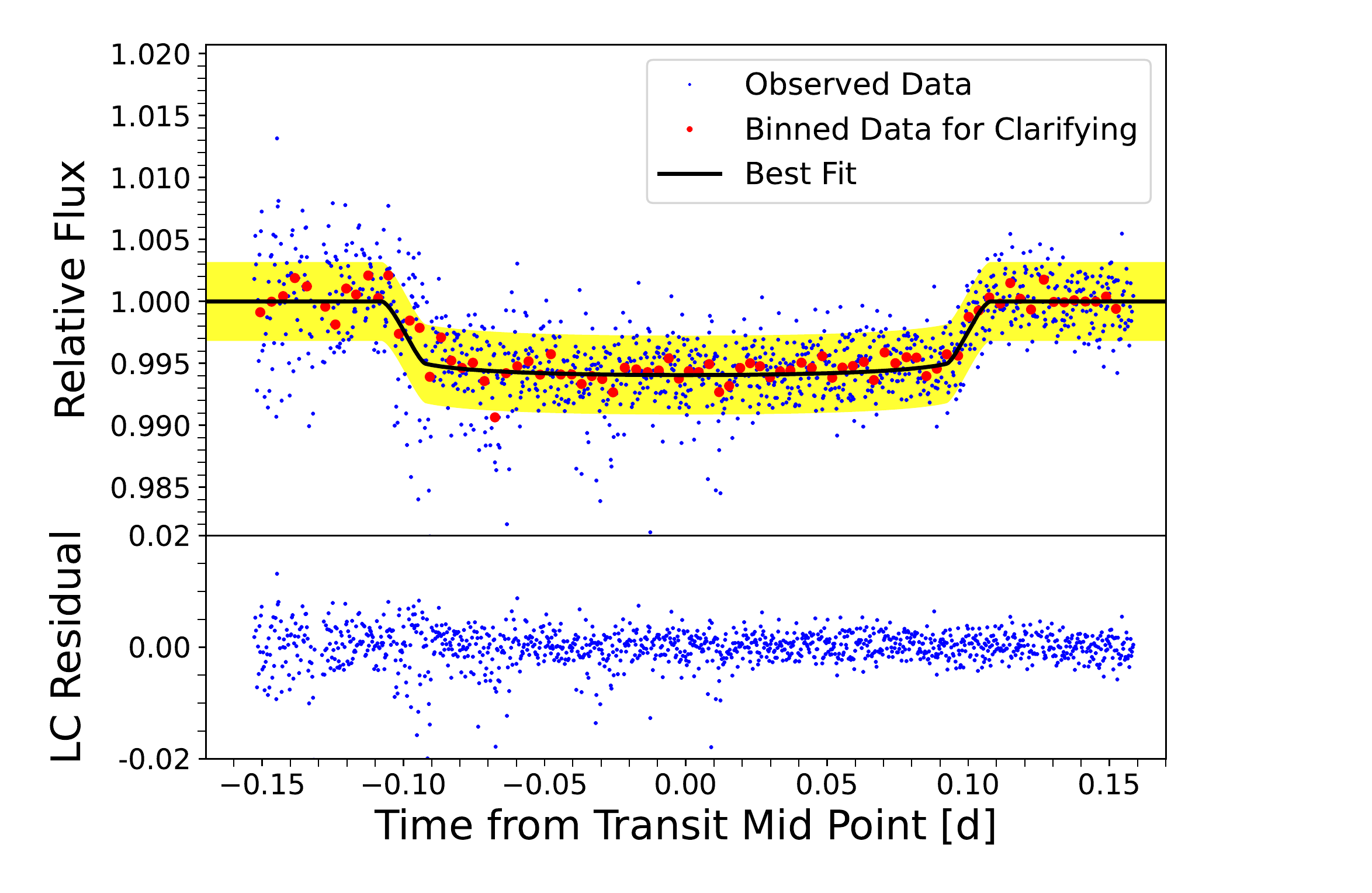}
   \includegraphics[width=3.38in]{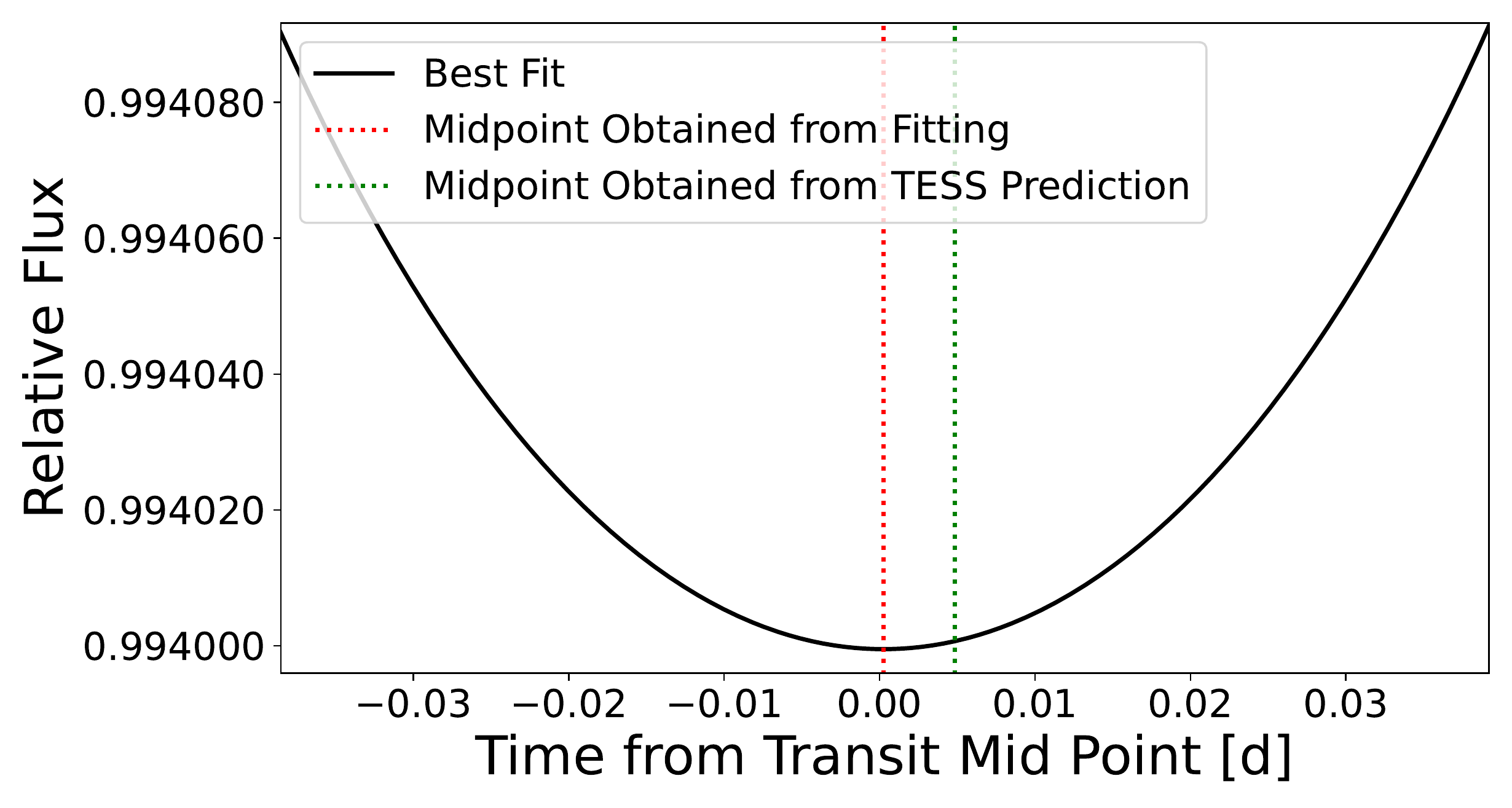}
  \caption{Fits to the archival SSO-Europa light curve from \citet{discoverpaper}. The symbols are similar to the top panel of Figure \ref{image: combinedfit}. The horizontal black line is the best fit transit to the data. The middle panel gives the LC residual from the fit where we can identify significant negative clumps. These clumps are likely due to the ground-based observation quality and are not correctable through transit modeling. The Shapiro-Wilk p-value is less than 0.01, indicating the residuals are not Gaussian. In comparison, the TESS light curve in Figure \ref{image: combinedfit} is well behaved. The bottom panel zooms in around the transit midpoint for clarity. The red vertical dotted line in the lower panel presents the fitting midpoint while the green vertical line indicates the midpoint predicted by TESS data with a constant period assumption. The offset is 6.97 minutes, equivalent to 4.40 $\sigma$. The TESS data indicate that the transit mid-point should be later in the SSO-Europa light curve which suggests that the orbital period of the planet may be decreasing with time. 
  }
 
\label{image: SSO-Europa} 
\end{figure}

We detrend the TESS light curve using the published ephemeris \citep{discoverpaper} as the baseline. We masked the data near the transit and fitted the data points with a linear function \citep{Yangatmos, YangLD}. High order polynomials (up to order 5) and cubic spline functions are also used to fit the trends in the data, but we find that the choice of function does not result in a significant difference to the light curve. We tested adding a trend (linear function of time) as a free parameter during transit modeling. The timing differences are within 0.2 minutes when compared to not including extra baseline parameters in light curve modeling. Our derived TESS light curve is shown in Figure \ref{image: combinedfit}.

\citet{discoverpaper} showed that the stellar variability does not appear to be significant enough to influence the transit modeling. The FFI data in 2019 shows the median brightness to be 9068$\pm$7\,$e^{-}\,s^{-1}$ while the 2021 10-minute cadence data shows the brightness to be 9070\,$\pm$11$e^{-}\,s^{-1}$. We note that the brightness comparison between different epochs is performed among the same kind of data product for consistency.
As a result, we do not find any significant evidence for variability in the TESS light curve either, i.e., any structure (other than transit) exceeding local median flux is less than 300 ppm.
We thus conclude that the impact of stellar variability on the derivation of transit parameters is negligible.

\subsection{Archival Data}
SSO-Europa light curve is a high-quality ground-based light curve with an out-of-transit baseline both before and after transit which is fully described in \citet{discoverpaper}. We detrend the data by masking the transit duration and fitting a linear function to the out-of-transit data points. We then 
fit the SSO-Europa photometry in a similar way to the process of fitting TESS light curves as described below. The fitted light curve is shown in Figure \ref{image: SSO-Europa}.

\subsection{Light Curve Fitting and Uncertainties}

The light curves we use are generated by our photometric pipeline, using 30-minute cadence FFI for TESS Sector 7 and 2-minute cadence TPF for TESS Sector 34. The detrended light curves are fitted with a transit model \citep{Mandel_Agol2002,Eastman2013} using a Monte-Carlo Markov Chain (MCMC) technique \citep{pymc,pya}. We choose a circular orbit for obtaining the transit mid-point, with the free parameters and priors the same as in previous work \citep{Yangatmos,YangLD}.  The MCMC chain takes 100,000 steps with the first 50,000 steps as burn-in. We derive transit parameters by fitting the folded light curve with data from each Sector. Each transit is then separately fitted for the specific transit mid-point. The priors when fitting the individual transit are set as the derived parameters from the folded light curve modeling.
The derived transit timings are listed in Table \ref{table: parameters}. As shown in the Table, we also tried a Keplerian orbit with eccentricity as a free parameter.

The Keplerian orbit model has as free parameters, the radius ratio of the planet to the host star ($R_{p}/R_{\ast}$), the orbital inclination ($i$), the semi-major axis (in the unit of stellar radii; $a/R_{\ast}$), the time of periapse passage, the longitude of the ascending node, the argument of periapsis ($\omega$), orbital eccentricity ($e$), and the limb darkening coefficients (LD; u1 for linear coefficient, u2 for quadratic coefficient). The parameterizations of $e$ and $\omega$ are $\sqrt{e}$sin$\omega$ and $\sqrt{e}$cos$\omega$, as the same of \citet{Eastman2013}. Except for the LD priors, the free parameters take uniform priors. The LDs assume Gaussian priors, centered at the stellar model prediction with a $\sigma$ of 0.05 \citep{TESSLD, YangLD}. Applying the stellar parameters from \citet{discoverpaper}, the priors of u1 and u2 are 0.31$\pm$0.05 and 0.22$\pm$0.05. The same steps are executed for circular orbit fitting. The 2-minute cadence TESS data provides the strongest constraints on LD parameters since it samples the transit ingress and egress the best. It yields a fitted $e$ of 0.28$\pm$0.14 (as shown in Table \ref{table: parameters}).

The fitting results and uncertainty are checked by simulating synthetic light curves from the data where the data points are randomly scattered
by their Gaussian uncertainty (the simulation details are similar to \cite{Yangatmos, YangLD}). We generate 1000 such simulated light curves
and assess the distribution of fit parameters with them.
The simulation indicates that most of the transit parameters have consistent parameter estimates between MCMC obtained values and the median results from the simulations. However, uncertainties of $e$ and a/R$_{\ast}$ are under-estimated by MCMC fitting by a factor of $\sim$ 2. MCMC fitting inevitably has difficulties in global parameter estimates when there are too many free parameters and some of which are not tightly constrained by the data \citep{Mackay2003, Hogg2018, YangLD}. We cite the uncertainties of $e$ and a/R$_{\ast}$ as the standard deviation of our simulation fit results, rather than the MCMC derived values, in this work.

We adopt double the timing uncertainty derived from the folded light curve, for the timing uncertainty of each epoch. We also provide timing uncertainties obtained from light curves of single visits (Table \ref{table: parameters}). For single visit light curve, we find that timing uncertainty inferred directly from MCMC fitting can be overestimated. The time series of the parameters in the chain present clumps when fitting single transits, indicating imperfect MCMC fits. Timing uncertainties derived from single-transit light curves are $\sim$ four times (for 30-minute data) and three times larger (for 2-minute data) than the result from folded light curves among one TESS Sector. According to error propagation, the uncertainty from the folded light curve should be 1/2 of the single visit given that every TESS sector contains four successful transits. Moreover, citing timing uncertainties from MCMC fitting to a single visit would lead to a significantly underestimated reduced $\chi^2$ of 0.13 when modeling timing evolution.

%%For the timing uncertainties for each epoch, we thereby adopt double the timing uncertainties derived from the folded light curves. 
%%twofold timing uncertainties from joint light curves as uncertainty for timing in each epoch. 
We note that the correlated noise is not included in our light curve modeling though a proper correlated noise model can potentially reduce the systematic errors \citep{PYCHEOPS, Patel2022}. The Gaussian process is commonly utilized when modeling the correlated noise \citep{Foreman-Mackey2017}. TESS data is stable enough that the noise dependency towards time (within 1 day) can be ignored in our precision of transit fitting \citep{TOIcatalog, IvshinaJosh}. TESS PDC light curves are modeled with correlated noise. We have compared our generated light curves and PDC light curves in previous work and find the difference is negligible in transit analyzing \citep{Yangatmos, YangLD, yangxo3b}. The timings of WASP-161b obtained by our generated light curve and PDC light curve are consistent within 0.1 minutes and the uncertainty from our generated light curve is 1.4 times larger.

%The uncertainties of single transits are cited for further analysis as half of the uncertainties fitted from the folded light curve in the relevant TESS sector. 
%We note that despite our simulations, it appears that the timing uncertainties are still slightly underestimated given that the best-fitted model of timings presents a reduced $\chi^2$ of 0.86 (as shown in Table \ref{table: parameters}). We also present uncertainties obtained from single visits for reference (as shown in Table \ref{table: parameters}).}

\subsection{Parameters Determined by a Combined-fit}

Some orbital parameters which are free parameters in light curve fitting are better constrained by the radial velocity (RV) data points. It, therefore, makes sense, if additional constraints on parameters can be obtained by undertaking joint fits to the RV and light curve data.
We constrain the value of $e$ by a combined-fit to TESS 2-minute cadence light curve and the radial velocity curve from \citet{discoverpaper}. The combined fit connects the constraints of both the LC and RV curves by adding the log-like likelihood of RV and LC models together \citep{pya}. The likelihood weights between the RV model and LC model are set equal. The combined model takes the same MCMC technique. The free parameter of the combined model includes all the free parameters of LC fitting and additional parameters for the RV curve, i.e., offset in radial velocity ($\gamma$), and the RV semi-amplitude ($K$). The fit gives an $e$ as 0.34$\pm$0.03. The fitting results of key parameters are as shown in Table \ref{table: parameters}.

Moreover, the MCMC fit to the RV curve alone using the Keplerian orbit model yields $e$ = 0.34$\pm$0.03 indicating that the RV data dominates the value of $e$ in the combined fit. Our derived $e$ is consistent with the 3 $\sigma$ upper limit of 0.43 reported by \citet{discoverpaper}.

We also attempt to fit a circular orbit model to the RV curve and find the fit is not as good as the eccentric Keplerian orbit model. 
The reduced $\chi^{2}$ of the circular model is $\sim$ 3, compared to the reduced $\chi^{2}$ of the Keplerian model $\sim$ 1. Applying the Bayesian Information Criterion \citep[BIC, details in][]{BICmostcited}, the $\Delta$BIC between the circular orbit and Keplerian orbit models is 27, preferring a Keplerian orbit model. 
The fitted light and radial velocity curves are present in Figure \ref{image: combinedfit}.

In addition, we jointly fit the light curves and radial velocities under the assumption of a circular orbit. 
We find that the derived timings from the joint fit are consistent (within 0.4$\sigma$) with timings from fitting the light curve only with a circular orbit. We note that the reduced $\chi^{2}$ of the RV curve is $\sim$ 3, and is indifferent to whether we do a joint fit or fit separately.

We conclude that the combined LC and RV fit with free eccentricity is the optimal model for the orbital parameters. A circular orbit is rejected by the RV curve. Fitting the LC and RV separately leads to inconsistency in the derived eccentricity (as shown in Table \ref{table: parameters}).

\section{Timing Variation and Interpretation}

\subsection{Timing Analysis}
\label{sec: timing}
 
The initial ephemeris is taken from \citet{discoverpaper} which gives a transit mid-time (T$_{c}$) at 2457416.5289$\pm$0.0011 days (Heliocentric Julian Date; HJD) and a period of 5.4060425$\pm$0.0000048 days. Table \ref{table: parameters} shows the transit mid-point in HJD, derived from the fits to the TESS and SSO-Europa light curves described earlier. The difference in minutes of the transit mid-point, with respect to the original ephemeris, is also shown in Table \ref{table: parameters}. The transit mid-point in SSO-Europa light curve is consistent at the 1.6$\sigma$ level with that in \citet{discoverpaper}. The uncertainty is estimated as the quadrature sum of the ephemeris uncertainty from \citet{discoverpaper} and timing uncertainty from SSO-Europa light curve. We note the previously reported ephemeris is indeed the best result given the uncertainties in the archival observations. However, the TESS data taken at later epochs, show that the difference in the transit mid-point becomes increasingly larger. We note that the timings from \citet{discoverpaper} are in HJD and TESS timings are in Barycentric Julian Date (BJD). The difference is within $\pm$4s which is negligible with the timing precision in our discussion.

\begin{table*}
\setlength{\tabcolsep}{2.5mm}
\begin{center}
\caption{WASP-161b parameters with 1 $\sigma$ significance region.}
\label{table: parameters}
\begin{tabular}{cccc}
  \hline
 \hline
\multicolumn{4}{c} {Transit Mid-points in the Data (HJD-2457000)} \\
\hline    
416.52890$\pm$0.00110$^{a}$ &  1124.71742$\pm$0.00083$^{b}$ &  1492.28605$\pm$0.00140(0.00265)$^{c}$ &  1497.69081$\pm$0.00140(0.00297)$^{c}$ \\
1508.50190$\pm$0.00140(0.00305)$^{c}$ &  1513.90827$\pm$0.00140(0.00335)$^{c}$ &  2232.81837$\pm$0.00094(0.00129)$^{c}$ &  2238.22514$\pm$0.00094(0.00128)$^{c}$  \\
2243.62942$\pm$0.00094(0.00132)$^{c}$ &  2249.03514$\pm$0.00094(0.00130)$^{c}$ \\
\hline
\multicolumn{4}{c} {Transit Mid-point Variation with Respect to Expected Ephemeris [minute] from \citep{discoverpaper}} \\
\hline   
 -0.00$\pm$1.58 &  -4.36$\pm$1.20 &  -65.25$\pm$2.01(3.81) &  -67.00$\pm$2.01(4.27) \\
 -68.52$\pm$2.01(4.39) &  -68.05$\pm$2.01(4.82) &  -202.77$\pm$1.35(1.86) &  -201.72$\pm$1.35(1.84) \\
 -204.26$\pm$1.35(1.90) &  --204.72$\pm$1.35(1.87) \\

\hline
\multicolumn{4}{c} {Transit parameters} \\
\hline   
Parameters  &  0 eccentricity      &  Free eccentricity    &  Combined LC and RV fit\\
            &                      &                     &Keplerian Orbit with Free Eccentricity \\
\hline
 R$_{p}$/R$_{\ast}$  & 0.0760$\pm0.0010$  & 0.0759$\pm$0.0008  & 0.0756$\pm$0.0007 \\
 i [degree] &  88.67$\pm$1.04   &    90.12$\pm$1.64 & 89.58$\pm$0.28\\
 a/R$_{\ast}$       & 8.83$\pm$0.36  & 6.62$\pm$1.16  & 6.57$\pm$0.45  \\
e &    0   &  0.28$\pm$0.14 &  0.34$\pm$0.04\\
u1 &   0.30$\pm$0.03   &   0.30$\pm$0.04  &   0.33$\pm$0.03\\
u2 &   0.22$\pm$0.04   &   0.21$\pm$0.04   &   0.19$\pm$0.03\\
std of residual (ppm)   & 2392  &   2392 &  2395  \\
reduced $\chi^{2}$   &  1.0002 &  1.0002  & 1.003  \\
 \hline
\multicolumn{4}{c} {RV model parameters} \\
 \hline
e        &         0                &       0.34$\pm$0.04        &   0.34$\pm$0.04  \\
$\gamma$ [km/s]    &  37.606$\pm$0.008     &       37.606$\pm$0.009              &   37.606$\pm$0.009  \\
$K$ [km/s]       &   0.257$\pm$0.014     &        0.254$\pm$0.022             &   0.255$\pm$0.036  \\
std of residual   & 0.0788  &   0.0447 & 0.0447  \\
reduced $\chi^{2}$   &  3.1053 &  1.0000  & 1.0000  \\
\hline 

\end{tabular}
\end{center}
\begin{flushleft}
Note. 
(a) Timing from \citet{discoverpaper}.
(b) Timing of SSO-Europa light curve.
(c) Timing of TESS light curves. The values inside parentheses refer to the uncertainties fitted from single visits.
\end{flushleft}
\end{table*}

We assess the accuracy of our TESS timing by applying our light curve generation and fitting analysis to WASP-58b and HAT-P-31b. The TESS timings are consistent within 1 $\sigma$ with respect to the published ephemeris \citep{Mallonn2019}. The targets for TESS timing check quoted here are selected randomly. In a separate work we discuss other comparisons of TESS timing precision\citep{shan2021}. 
We also note that TESS timing has been used for reliable long-term monitoring of WASP-12b \citep{2021wasp12bTESS} and to detect transit timing variations in WASP-4b and XO-3b \citep{WASP-4b,yangxo3b}.

\begin{figure}
  \centering
   \includegraphics[width=3.5in]{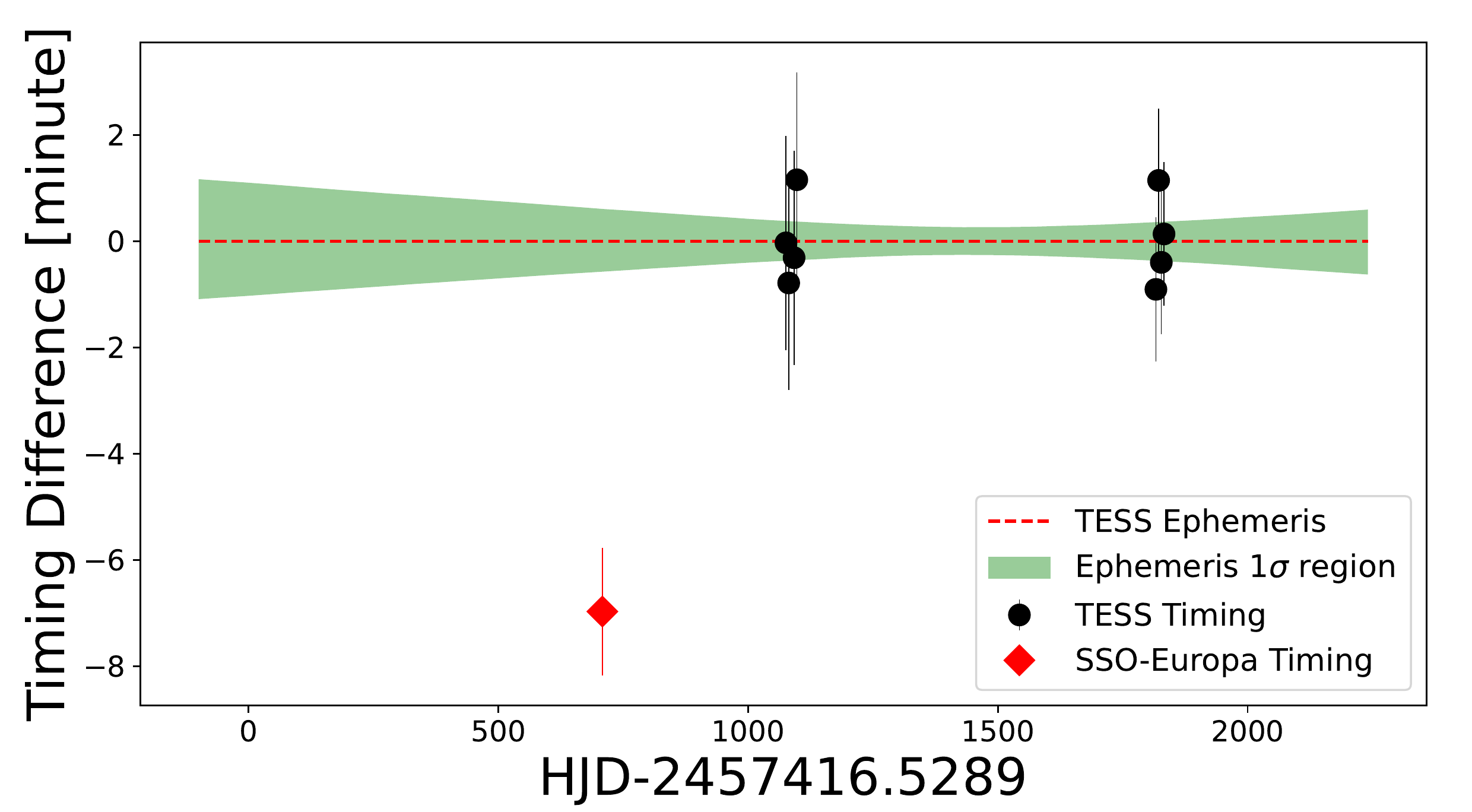}
  \caption{Our derived difference in transit midpoints of WASP-161b from SSO-Europa and TESS data. The
  ephemeris is obtained from fitting to only TESS timings (black points), using the MCMC method. The green area indicates the 16$\%$ to 84$\%$ percentile region of all the MCMC chain predictions. The red point is the transit midpoint difference in the archival SSO-Europa data compared to the TESS ephemeris. The mid point of the SSO-Europa transit is 6.6 minutes earlier than that estimated from the TESS data.}
  \label{image:td}
  \end{figure}

We find that if the TESS timings are accurate, the transit midpoint of SSO-Europa light curve is earlier by 6.97 minutes relative to the constant period assumption (as shown in Figure \ref{image: SSO-Europa} and Figure \ref{image:td}). This is at a significance level of 4.63 $\sigma$ calculated as the sum in quadrature of the timing uncertainty from SSO-Europa light curve and TESS ephemeris uncertainty (as shown in Figure \ref{image:td}). The TESS ephemeris uncertainty is evaluated as the 1$\sigma$ region of the MCMC chain from a linear fit to the TESS timings.

\begin{figure*}
\centering
\includegraphics[width=7.5in]{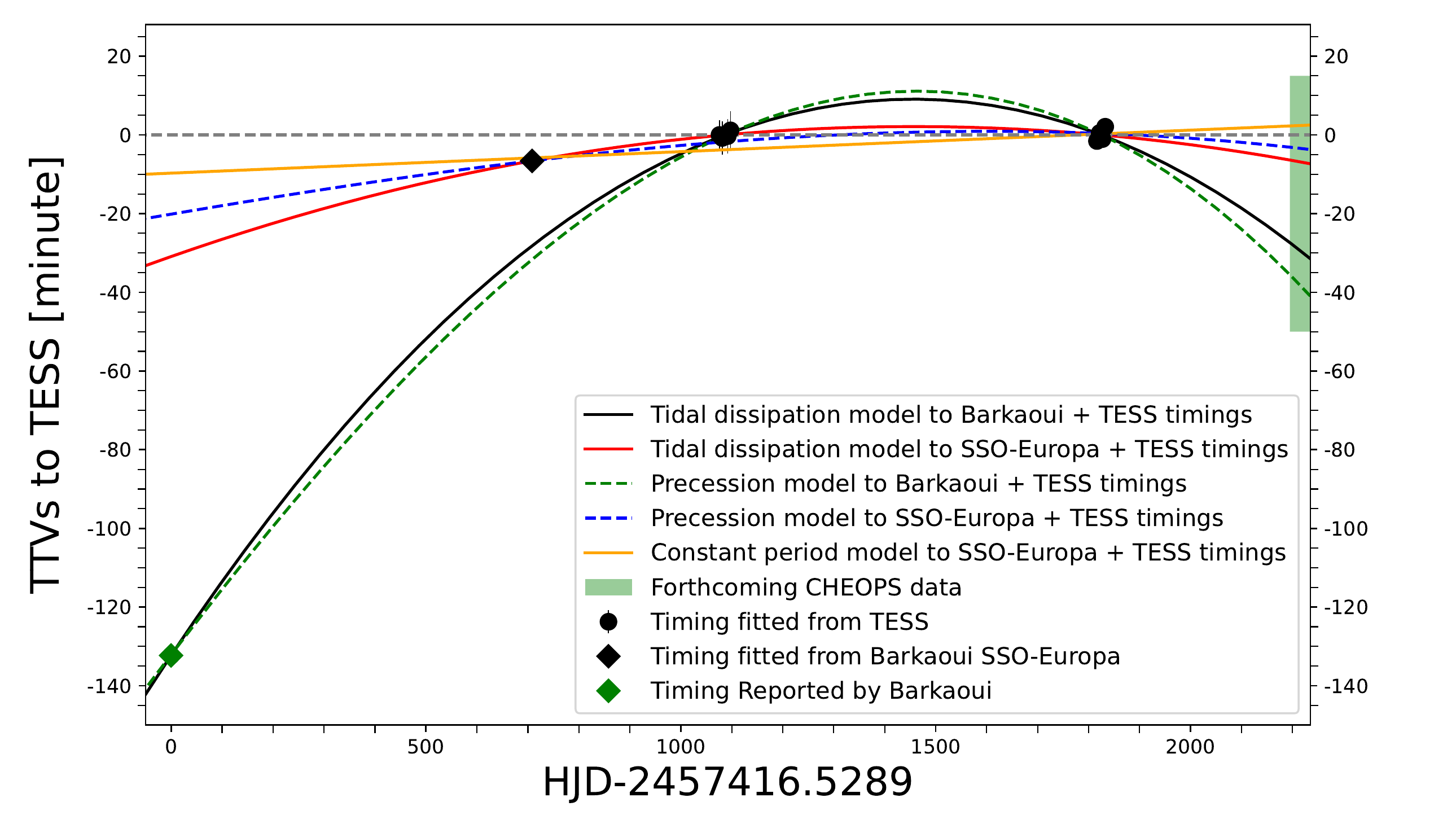}
\caption{
Figure showing the evolution in the WASP-161b transit midpoint as a function of time. We show the difference in the transit mid-points relative to the TESS-measured ephemeris. 
The black points indicate the timing obtained from the SSO-Europa and TESS observations. 
The orange solid line presents the linear model with a constant period fit to the TESS observations and our timings derived from the archival SSO-Europa observations. 
The red solid line is the quadratic model (period changing linearly with time as expected for tidal dissipation) fit to the timings from SSO-Europa and TESS light curves. The black solid line is the quadratic model to the \citet{discoverpaper} reported and TESS timings which is inconsistent with the timing from SSO-Europa light curve. 
The dashed green blue and green lines indicate precession models which are not significantly different from quadratic models when fitting the timing evolution. The green region indicates the forthcoming CHEOPS data which will be able to distinguish between the models.}
\label{image: models}
\end{figure*}

Conversely, the timing from \citet{discoverpaper} yields a 132.30 minutes deviation at 53.03 $\sigma$ when compared to TESS timings. TESS observations indicate transit midpoints which are earlier by $\sim$ 67.24$\pm$2.22 minutes at Jan. 2019, and $\sim$ 202.71$\pm$0.91 minutes at Jan. 2021, compared to the ephemeris from \citet{discoverpaper}. The values are the mean values among the four transits in each period of TESS observation, each of which are shown in Table \ref{table: parameters}. The uncertainty is the mean of the quadrature sum of the timing uncertainties.

We can not distinguish whether the timing from SSO-Europa light curve or the timing from \citet{discoverpaper} is more accurate because of the constant period assumption in their work and the absence of the timings from the individual epochs in \citet{discoverpaper}. We note that their timings are consistent with our analysis of the timing from SSO-Europa light curve, under the assumption of a constant period, which however, seems unlikely.

Furthermore, we can conclude that the transit mid-point in the SSO-Europa data is statistically different compared to the transit mid-point in the TESS data (Figure \ref{image:td}) and similarly, that the TESS data show transit mid-points which are different from the published ephemeris of WASP-161b by large amounts.

We fit the evolution in transit midpoints with two scenarios: a period evolution $\dot{P}$ which is zero and a $\dot{P}$ which is constant. We refer to these as the linear model and the quadratic model respectively.
We find that a linear model is not as good as a quadratic model when explaining the transit mid-points from both the SSO-Europa and TESS observations (as shown in Figure \ref{image: models}). The BIC of the quadratic model is 10.33 smaller than the linear model based on our analysis of the archival data. Furthermore, if we use the ephemeris from the \citet{discoverpaper}, we find that the BIC of the quadratic model is 6,500 smaller than any linear model. The quality of fits are shown in Table\ref{table: models}.

\begin{table*}
\setlength{\tabcolsep}{3mm}
\begin{center}
\caption{Models for Timing Deviation}
\label{table: models}
\begin{tabular}{cccc}
  \hline
 \hline
\multicolumn{4}{c} {Linear Model: t$_{d}$ = a$_{l}$$\times$t + b$_{l}$} \\
\multicolumn{4}{c} {Period Decaying Model: t$_{d}$ = a$_{d}$$\times$t$^{2}$ + b$_{d}$$\times$t + c$_{d}$} \\
\multicolumn{4}{c} {Precession Model: t$_{d}$ = a$_{p}$$^{1}$$\times$cos($\omega_{t0}$ + $\omega_{t}$$\times$t) + b$_{p}$$\times$t + c$_{p}$}\\
\hline   

Parameters  &  Only TESS Timing      &  SSO-Europa + TESS Timing   & \citet{discoverpaper} reported + TESS Timing\\
 \hline
                    \multicolumn{4}{c}   { Linear Model }\\
\hline 
 
a$_{l}$ &   -0.1836$\pm$1.5$\times$10$^{-6}$  &   -0.1796$\pm$3.1$\times$ 10$^{-6}$   &   -0.1260$\pm$0.0003\\
b$_{l}$ &   132.3$\pm$3.5   &   125.7$\pm$5.9   &  43.1$\pm$550.2\\
%reduced $\chi^{2}$   &  -- &  3.54  & 1569.83 \\
%\textbf{BIC}   & -- &  36.30  & 14132.86  \\
%reduced $\chi^{2}$   &  -- &  2.07  & 929.76 \\
%\textbf{BIC}   & -- &  18.92  & 6512.73  \\
reduced $\chi^{2}$   &  -- &  2.03  & 940.93 \\
BIC   & -- &  18.67  & 6590.90  \\
 \hline
\multicolumn{4}{c} {Period Decaying Model} \\
\hline 

a$_{d}$      &   --     &   -1.54$\times$10$^{-5}$$\pm$3.0$\times$10$^{-6}$    &   -6.67$\times$10$^{-5}$$\pm$1.3$\times$10$^{-6}$  \\
b$_{d}$      &   --     &   -0.1387$\pm$0.0080   &    0.0106$\pm$0.0027                \\
c$_{d}$      &   --     &   101.7$\pm$4.8        &     0.0$\pm$1.4               \\
$\dot{P}$ (days per day)     &   --     &   -1.16$\times$10$^{-7}$$\pm$2.2$\times$10$^{-8}$    &   -5.01$\times$10$^{-7}$$\pm$9$\times$10$^{-9}$ \\
%(second per orbit per second)
%reduced $\chi^{2}$   &  -- &  1.000  &  1.171 \\
%\textbf{BIC}  & -- &  15.59  & 17.13  \\
%reduced $\chi^{2}$   &  -- &  0.43  &  0.53 \\
%\textbf{BIC}  & -- &  9.16  & 9.79  \\
reduced $\chi^{2}$   &  -- &  0.29  &  0.33 \\
BIC  & -- &  8.34  & 8.58  \\
\hline 
\multicolumn{4}{c} {Precession Model} \\
\hline
a$_{p}$      &   --     &   -842.49$\pm$0.57    &   -842.50$\pm$0.57  \\
$\omega_{t0}$  &   --     &   1.689$\pm$0.008    &   2.114$\pm$0.043 \\
$\omega_{t}$      &   --     &   3.22$\times$10$^{-4}$$\pm$1.6$\times$10$^{-5}$    &  4.42$\times$10$^{-4}$$\pm$1.0$\times$10$^{-5}$ \\
b$_{p}$      &   --     &  -0.409$\pm$0.008    & -0.323$\pm$0.014   \\
c$_{p}$      &   --     &   0.12$\pm$5.78   &    -435.50$\pm$31.79  \\
$\frac{d\omega}{dN}$ (rad)     &   --     &   1.74$\times$10$^{-3}$$\pm$0.08$\times$10$^{-3}$    &   2.39$\times$10$^{-3}$$\pm$0.06$\times$10$^{-3}$  \\
%reduced $\chi^{2}$   & -- &  1.092  & 1.102  \\
%\textbf{BIC}   & -- &  18.46  & 18.59  \\
%reduced $\chi^{2}$   & -- &  0.54  & 0.72  \\
%\textbf{BIC}   & -- &  11.47  & 12.40  \\
reduced $\chi^{2}$   & -- & 0.36  & 0.46  \\
BIC   & -- &  10.61  & 11.10  \\
\hline 
\end{tabular}
\end{center}
\begin{flushleft}
Note. 
(1) a$_{p}$ has a uniform prior of (-843.5, -841.5), according to Equation \ref{equ:precession}.
\end{flushleft}
\end{table*}

\subsection{Possible Interpretation of the Measurements}
\label{models}

The origin of orbital period decay in planet-satellite systems and hot Jupiters have been discussed in the literature \citep[][and references therein]{MurrayDermott, CorreiaLaskar, Goldreich1966, Ragozzine2009,Hansen2010, Hansen2012, 2020wasp12b}.
Gravitational tides circularize elliptical orbits on timescales of $\sim$1\,Myr unless the eccentricity is excited by the gravitational influence of an outer planetary companion \citep{circularization}. 
For tides to cause a decay in the orbital period of the planet, the star must be rotating slower than the orbital velocity of the planet \citep{Hut1981,MurrayDermott, Themodel,Millholland2018}. In the case of WASP-161b, \citet{discoverpaper} has measured the stellar rotation to be faster than the orbital velocity, with a rotation period of $\sim$4.8 days compared to an orbital period of $\sim$5.4 days. Thus, it seems unlikely that tidal dissipation can explain the TTVs unless the planet is in a retrograde orbit.

Alternately, the rotational and tidal bulges on the planet can introduce apsidal precession; this is however thought to affect the light curve shape to a greater degree than transit timings \citep{Pal2008, Ragozzine2009}. We do not detect any clear evidence for variation in the transit duration. %although this may partly be due to the limited signal to noise in the SSO-Europa data. 

We plot our transit mid-point variations and compare them to those expected from the
period decay and precession models in Figure \ref{image: models}.

The period decay model yield transit times ($t_\mathrm{tra}$)for the Nth transit relative to the transit zero point $t_{0}$ of \citep{2017wasp12b,2020wasp12b}:  
\begin{equation}
\begin{aligned}
t_\mathrm{tra}(N) &= t_0 + NP + \frac{1}{2}\frac{dP}{dN}N^2\\
\end{aligned}
\end{equation}
where $t_\mathrm{tra}(N)$ is transit timing at Nth transit relative to the timing zero point $t_0$, $P$ is the period at $t_0$. $t_d$ is the difference in transit time relative to that expected from a constant orbital period i.e. $t_d=t_{tra}(N)-t_{0}-NP$.
The timing variance is thus a quadratic function (as shown in Table \ref{table: models}), and leads to a $\dot{P}$ of -1.16$\times$10$^{-7}$$\pm$2.25$\times$10$^{-8}$ days per day (or $-3.65$ s/year). 
The $P$/$\dot{P}$ is 0.128 Myr. 
Given the derived eccentricity of WASP-161b, and the expected circularization of the orbit on relatively short timescales, a companion in an outer orbit is needed to excite and maintain the high eccentricity \citep{Naoz2011,Dawson2018}. \citet{Yu2021} present a non-linear interaction, very effective in tidal dissipation, enabling a dissipation timescale as short as 10$^4$ yr. 

For apsidal precession, the transit timing is predicted as \citep{2017wasp12b,2020wasp12b}:
\begin{equation}
\label{equ:precession}
\begin{aligned}
t_\mathrm{tra}(N) &= t_0 + N P_s - \frac{e P_a}{\pi} \cos{\omega(N)}\\
\omega(N) &= \omega_0 + \frac{d\omega}{dN} N \\
P_s &= P_a \left(1 - \frac{1}{2\pi}\frac{d\omega}{dN}\right),
\end{aligned}
\end{equation}
where $P_s$ is the sidereal period, $P_a$ is the anomalistic period and $\omega(N)$ is the precession amplitude of the Nth orbit. As shown by
\citet{Ragozzine2009}, the precession amplitude is proportional to the Love numbers of the planet and the star.
%For WASP-161b, the influences of planet and star are comparable that tidal precession rate ($\dot{\omega}_{\rm tidal}$) should be expressed as \citep{Ragozzine2009}:
%\begin{equation}
%\label{equa: precession}
%\begin{aligned}
%\dot{\omega}_{\rm tidal} = &
%\frac{15}{2} k_{2*} \left( \frac{R_*}{a}\right)^5 \frac{M_p}{M_*} f_2(e) n + \\
%&\frac{15}{2} k_{2p} \left( \frac{R_p}{a}\right)^5 \frac{M_*}{M_p} f_2(e) n,
%\end{aligned}
%\end{equation}
%where $f_2(e)$ is expressed as: 
%\begin{eqnarray*}
%f_2(e) & = & (1-e^2)^{-5}(1 + \frac{3}{2} e^2 + \frac{1}{8} e^4).
%\label{f2e}
%\end{eqnarray*}
%Since $e$ is not negligible for WASP-161, the $f_2(e)$ is not 1. $k_{2*}$ and $k_{2p}$ are the star and planet's Love numbers.

The observational precession rate is derived from TTVs fitting. The TTVs are fitted with a combined-formula including a cosinusoidal and a linear formula (as shown in Table \ref{table: models}). We use the best fitting TTVs from our generated SSO-Europa and TESS light curves. However, we find that the \citet{discoverpaper} reported values and TESS timing give a similar result (within 30$\%$). We obtain a precession rate of 3.22$\times$10$^{-4}$$\pm$1.6$\times$10$^{-5}$ rad per day (or $\sim$6 deg/yr), corresponding to a cycle of 53$_{-2.5}^{+2.7}$ years. 

Both models provide reasonable fits to the observational TTVs, though the precession model has more free parameters (as shown in Table \ref{table: models}) and results in abnormal Love Numbers of 141 and 123 for the planet and star respectively. The tidal model clearly also has issues given the rotation rate of the star. Further data is clearly required to understand the unusual orbital parameters of WASP-161b.
%Thus, in the absence of additional data, we favor the decaying period scenario to explain the measurements of WASP-161b. 

%\begin{figure}
%  \centering
%  \includegraphics[width=3in]{migration_calibration1.pdf}
%  \caption{WASP-161b timing variation, compared to linear best fits: Top panel shows the \textbf{timing variations} assuming the fit to the \citet{discoverpaper} ephemeris and TESS timings. The bottom panel shows the offset assuming a fit to SSO-Europa and TESS timings. The symbols are the same as Figure \ref{image: models}. As discussed in the text, we think the lower panel is a more accurate representation of the measurements.}
% \label{image: comparison}
%\end{figure}

\subsection{The Need for More Observations}

We have presented the evidence for TTVs of WASP-161b and discussed plausible interpretations of the measurements. It is interesting to note that of the four observed planets (WASP-161b, WASP-12b, WASP-4b, XO-3b) with TTVs, all have earlier timing than the constant period model. Apsidal precession should result in an unsigned timing variation which is not the case. Thus, the declining period model with the presence of an undetected outer planet to explain the significant eccentricity, is the preferred explanation for the TTVs in WASP-161b.
We note that the magnitude of the transit time difference in SSO-Europa light curve is well matched to the difference between HJD and JD on the date of the SSO-Europa observations; although \citet{discoverpaper} states SSO-Europa light curve is in HJD in their data release notes
\footnote{\url{https://cdsarc.cds.unistra.fr/viz-bin/cat/J/AJ/157/43##/browse
}}\footnote{Communication between us and the WASP team (Coel Hellier), has indicated that the timing in \citep{discoverpaper} is in Coordinated Universal Time (UTC) rather than TDB (Barycentric Dynamical Time). This would imply that the timing difference (SSO-Europa) compared to our estimates in Table \ref{table: parameters} would be reduced by 1.135 minutes.}. 
%\FanComments{Moreover, TESS quotes times as Barycentric Dynamical Time (TDB). We note that it would induce a 1.135 minute earlier timing if SSO-Europa light curve was in Coordinated Universal Time (UTC).}
%the WASP documentation indicates that the light curves are in HJD \citep{wasp2010}. 
%confirmation of the timing trend will have to await future observations.

Further observations with more precise timing are needed to distinguish between the different scenarios presented. We have been approved for observations with the space telescope CHEOPS \citep{Benz} for two transits around Jan. 2022.

The transit mid-points predicted from the different models can be distinguished with the CHEOPS observation \citep{Benz}, given the quoted precision of CHEOPS photometry and sampling rate ($\sim$ 700 ppm precision with an exposure time of 30s). The CHEOPS observation would present a TTV larger than 200 minutes when compared to the ephemeris from \citet{discoverpaper} if the period decay model is as derived from the \citet{discoverpaper} and TESS timings (Figure \ref{image: models}). In the favored scenario, the period decay model obtained from SSO-Europa and TESS timings would yield a TTV of $\sim$ 5 minutes when compared to the constant period timing predicted from TESS observations. The difference between a period decay model fit to the \citet{discoverpaper} ephemeris and TESS data would be 22 minutes relative to the model fit to the SSO-Europa and TESS data. Thus, we conclude that forthcoming observations will be able to enhance our understanding of the orbital evolution of WASP-161b.

\section{Summary}

In this work, we report the discovery of TTVs of WASP-161b which presents a transit mid-point difference of 67 and 203 minutes in TESS 2019 and 2021 observations, compared to the ephemeris from \citet{discoverpaper}. We reassess the timing in the archival 2018 SSO-Europa data and find that it is consistent with the \citet{discoverpaper} timing assuming a constant period. However, we find that the transit mid-point from SSO-Europa light curve presents a 6.97-minute difference compared to ephemeris estimated from the TESS data under a constant orbit period assumption.

%The most likely explanation for the TTVs is tidal dissipation which predicts 
The TTVs appear to follow a constant period derivative. The $\dot{P}$ is -1.16$\times$10$^{-7}$$\pm$2.25$\times$10$^{-8}$ days per day when fitting to the SSO-Europa and TESS timings. 
%The corresponding $Q_p'$ and $Q_*'$ are 2.4$\times$$10^{5}$$\pm0.3$$\times$$10^{5}$ and 2.3$\times$$10^{4}$$\pm$$0.3\times10^{4}$. 
The $\dot{P}$ obtained from taking the \citet{discoverpaper} ephemeris as is, and TESS timings are $\sim$ 4 times larger - we think this scenario is unlikely because of the measurement uncertainties in \citet{discoverpaper}. However, distinguishing between the two scenarios needs further observations.

We have been approved CHEOPS observation in Jan. 2022 for further observations of WASP-161b. CHEOPS high precision photometry and sampling rate will enable us to obtain good timing estimation which is crucial for further constraining the period decay rate. Combining archival radial velocity data with the TESS measurements suggests that WASP-161b has a high eccentricity orbit, a period of 5.41 days, and is undergoing a period decrease. These properties are different from the previously confirmed hot Jupiter WASP-12b which has a shorter period but with an orbital decay timescale that is 20 times smaller. The hot Jupiters which show robust evidence for TTVs seem to be presenting evidence for the orbital period to be decreasing, in some cases with significant eccentricity and in other cases with zero eccentricity. They may therefore be important laboratories for investigating the origin of hot Jupiters.

\section{acknowledgements}
This work made use of the NASA Exoplanet Archive \citep{ExoplanetArchive} 10.26133/NEA12 table and PyAstronomy\footnote{https://github.com/sczesla/PyAstronomy}. We would like Xing Wei, Su-Su Shan for helpful discussion. Fan Yang 
acknowledges funding from 
the Cultivation Project for LAMOST Scientific Payoff and Research Achievement of CAMS-CAS, the National Key Research and Development Program of China (No. 2016YFA0400800), the National Natural Science Foundation of China (NSFC) (No. 11988101, No. 11872246), and the Beijing Natural Science Foundation (No. 1202015).

\bibliographystyle{aasjournal}
\bibliography{ref}
\end{document}